\UseRawInputEncoding
\documentclass{article}

\usepackage{PRIMEarxiv}

\usepackage[utf8]{inputenc} 
\usepackage[T1]{fontenc}    
\usepackage{hyperref}       
\usepackage{url}            
\usepackage{booktabs}       
\usepackage{amsfonts}       
\usepackage{nicefrac}       
\usepackage{microtype}      
\usepackage{lipsum}
\usepackage{fancyhdr}       
\usepackage{graphicx}       
\graphicspath{{media/}}     
\usepackage{amsmath}

\usepackage[utf8]{inputenc}
\usepackage{textgreek} 
\usepackage{listings}

\title{Systems Engineering for Autonomous Vehicles; Supervising AI using Large Language Models (SSuperLLM)}

\author{
  Diomidis Katzourakis \\
  Autonomous Systems, Vehicle Dynamics and Controls Engineering Services\\
  \texttt{diomkatz@gmail.com} \\
}

\begin{document}
\maketitle

\begin{abstract}
Generative Artificial Intelligence (GAI) and the idea to use hierarchical models has been around for some years now. GAI has proved to be an extremely useful tool for Autonomous Vehicles (AVs). AVs need to perform robustly in their environment. Thus the AV behavior and short-term trajectory planning needs to be: a) designed and architected using safeguarding and supervisory systems and b) verified using proper Systems Engineering (SysEng) Principles. Can AV Systems Engineering also use Large Language Models (LLM) to help Autonomous vehicles (AV) development? This reader-friendly paper advocates the use of LLMs in 1) requirements (Reqs) development and 2) Reqs verification and 3) provides a proof-of-concept of AV supervisory control. The latter uses a simulation environment of a simple planar (bicycle) vehicle dynamics model and a Linear Quadratic Regulator (LQR) control with an LLM Application Interface (API). The Open-Source simulation SW is available from the author \footnote{The author currently works as Platform Requirements and Verification Lead at Waabi.ai (joined 2024). This article derives from work that the author did before joining Waabi.ai working as independent developer. Before that he worked as SysEng Manager on Motion Control and Vehicle Dynamics at Waymo (2019-2023, CA, USA), Product Design and Control Systems for Autonomous Systems at Apple Inc (2015-2019, CA, USA), CAE Active Safety assignment leader at Volvo Cars (2012-2015, Gothenburg, Sweden).} accessible \footnote{ \url{https://colab.research.google.com/drive/1tRLu1l-bK-zClOGACPyK3Oyzf7mtSxKb?usp=sharing} \\under the MIT license. Users of the SW need to use their own Google API key "GOOGLE\_API\_KEY='' as depicted in the SW.} to the readers so that they can engage into the AV stack, LLM API and rules, SysEng and Reqs and fundamental vehicle dynamics and control. 
\end{abstract}


\section*{Abbreviations}

\begin{minipage}{0.45\textwidth}
    \begin{itemize}
        \item \textbf{AI}: Artificial Intelligence
        \item \textbf{AV}: Autonomous Vehicles
        \item \textbf{LLM}: Large Language Model
        \item \textbf{SW}: Software
        \item \textbf{GAI}: Generative AI
    \end{itemize}
\end{minipage}%
\hfill
\begin{minipage}{0.45\textwidth}
    \begin{itemize}
        \item \textbf{ML}: Machine Learning
        \item \textbf{Reqs}: Requirements
        \item \textbf{VnV}: Verification and Validation
        \item \textbf{API}: Application Interface
        \item \textbf{SysEng}: Systems Engineering
    \end{itemize}
\end{minipage}

\section{Introduction}
\label{Introduction}

The development of AV requires a big knowledge set, from perception and behavior prediction to rules for legal compliance and varying fidelity maps (high definition or not depending on the application). Ground up development of this knowledge is truly a time and cost consuming endeavor. Waymo is the accepted leader in AVs development/realization, offering a true service. Still it has taken Waymo more than 20 million miles (c.f. \cite{1}) of test driving and quite some years to achieve its feat. As an example, refer to \cite{2} where any possible conflict (collision) behavior has to be a) identified, b) described, c) set up as goal and/orReq and d) verified. It is labor intensive and requires some “expensive” thinking! AVs rely to (and have to) the possible extent on Machine Learning (ML) since it can perform tasks without explicit instructions throughout the autonomy stack. 

\begin{figure}[ht]
 \centering
 \includegraphics[width=\linewidth]{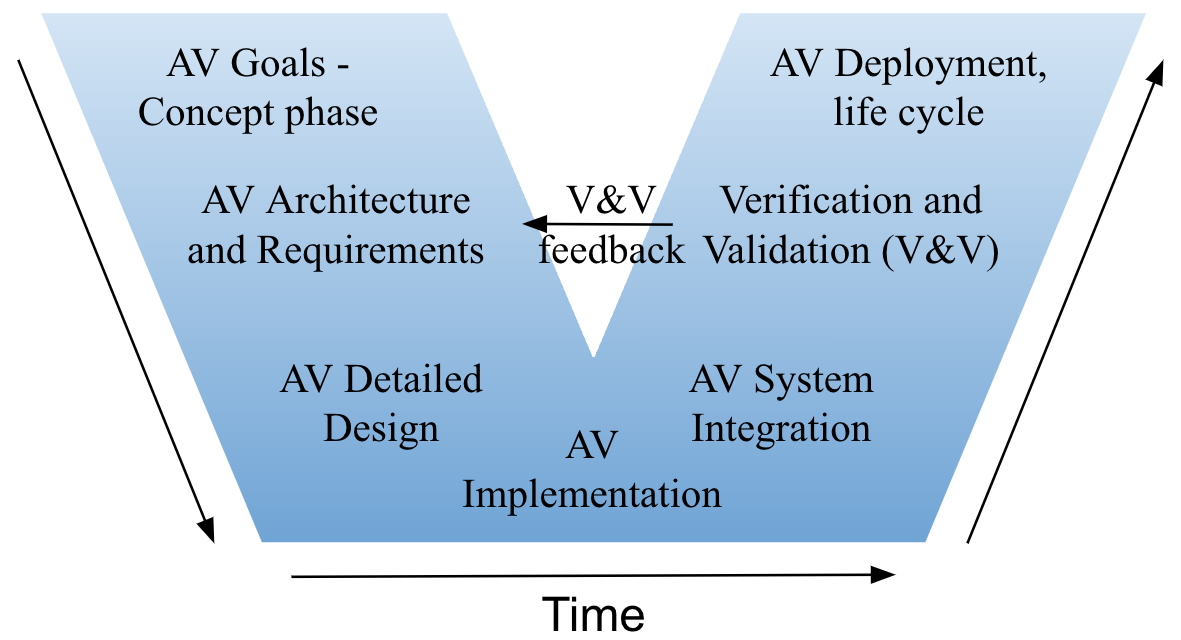}
 \caption{The V-model of the systems engineering process \cite{0}. VnV feedbacks back to architecture and Reqs.}
 \label{fig:Fig_0}
\end{figure}

Generative AI (GAI) and the idea to use hierarchical models (to generalize and train policies mimicking expert behaviors) have been around for some years now \cite{2} and has been an extremely useful tool for AVs \cite{4}. As \cite{5} states “Predicting the future behavior of road users is one of the most challenging and important problems in autonomous driving.” It is necessary “to forecast the short-term movements and long-term trajectories of vehicles” \cite{6}. \cite{7} states the importance of the AV to “perform robustly in safety-critical settings,” thus the ML methodology for EGO BEHAVIOR and PLANNING needs to be: a) designed and architected using {\em safeguarding and supervisory} systems and b) verified using authentic Systems Engineering (SysEng) Principles. Refer to Fig. \ref{fig:Fig_0} for the typical SysEng V-model. The main goal of SysEng is to {\it fix the defect in the concept phase}, c.f. \cite{8}) and be {\it "Safe by Design"} \cite{26}. This enabled by Decomposing the System target-goals to principled requirements (Reqs). The idea is that if all the children Reqs are met, then also the parent Req is met (c.f. \cite{0}). Devising the correct functional Reqs and Verification and Validation (VnV) strategy is a key component to success (c.f. Section \ref{FuncyD}). The successful execution of VnV and as consequence the safety of ML dominated Cyber-physical systems (i.e. AVs) is a research field itself \cite{25,27}. 

\subsection{Hypothesis}
\label{Hypothesis}

Can AV Systems Engineering also use Large Language Models (LLM) to help Autonomous vehicles (AV) development (i.e. \cite{9})? We believe yes and some recent studies and their findings support our hypothesis (i.e. \cite{18,19,20,21}). The current paper prototypes our Supervisory control methodology and validates the hypothesis. 

\vspace*{-0.05in}
\section{State of the art}
\label{SOFA}
\vspace*{-0.05in}

In 2014 SAE \cite{10} published J3016 the SAE Levels of Driving Automation™ in the context of motor vehicles and their operation on roadways. It defines Level 0 (no driving automation), to L2 (supervised limited “autonomy”, which is what most automotive manufacturers (OEMs) offer nowadays), Level 4 (only a few AV companies offer) and Level 5 (full driving automation). Financial competition in industrial road AVs sparked an autonomous “race.” 
 
Most of the big OEMs and (probably) hundreds of start-ups around the globe started the race, but they had to slow down or stop due to the significant complexity of the problem and the associated cost-resources needed to solve it. The contenders still in the race are Mobileye, that designed its system intelligently and has been a true commercial success used in tens of million vehicles, and, of course, the groundbreaking Tesla, with its Autopilot system. Nvidia has built an end-to-end, software-defined platform to develop AVs at scale with a number of partnerships with prestigious OEMs. Chinese Pony.ai and Baidu have been working towards L4 autonomy, with the latter receiving in December 2022 the first license to test the AV with no safety operator in Beijing. Zoox, a subsidiary of Amazon, is trying to deploy an autonomous robotaxi service. The US GM Cruise began offering taxi services via entirely autonomous means in 2021, suspending the service in October 2023. The accepted leader is Alphabet owned Waymo, which has a true ride-hail service that is already available to the public in Phoenix (Arizona), San Francisco and Los Angeles (and expanding), based on a mature System architecture, designed and verified with SysEng principles. 

\begin{figure}[ht]
 \centering
 \includegraphics[width=\linewidth]{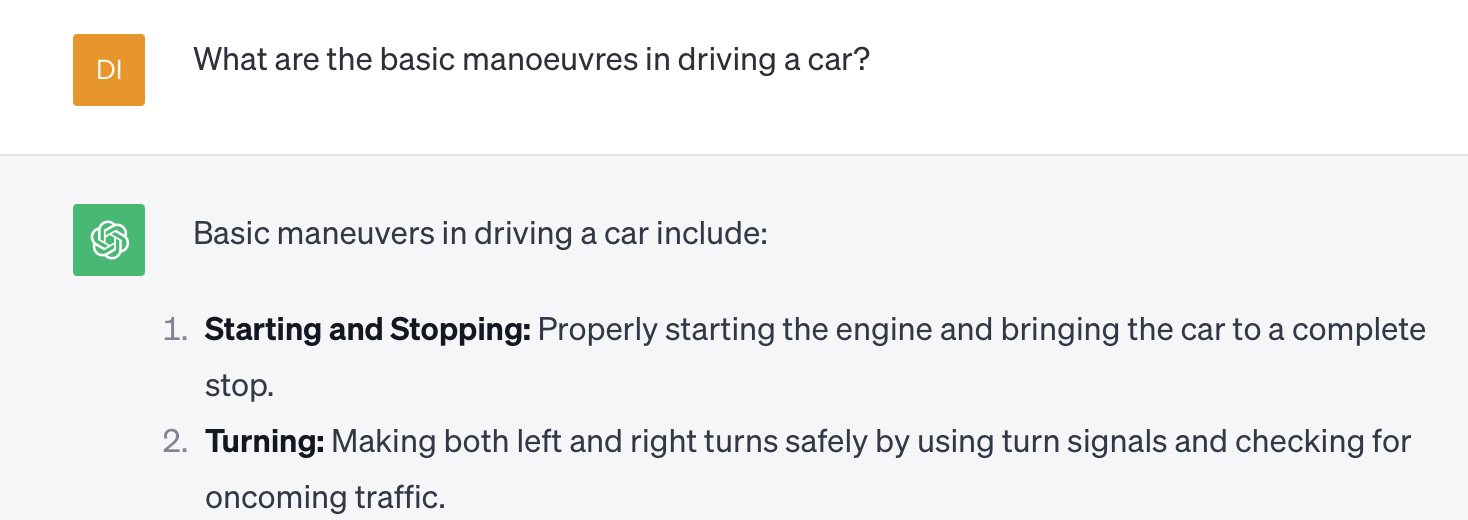} 
 \caption{Part of the answer of ChatGPT-3.5 in the question “What are the basic manoeuvres in driving a car?”}
 \label{fig:Fig_1}
\end{figure}

\section{Methodology}
\label{Methodology}

\vspace*{-0.05in}
\subsection{Requirements Development}
\label{ReqsDev}
\vspace*{-0.05in}

An LLM, in the context of artificial intelligence and natural language processing, refers to a neural network-based model that has been trained on text data to process and generate human language. Some well-known examples of large language models include ChatGPT, Gemini (Google), etc. each with its own architecture. These models have found applications in a wide range of fields, from chatbots and virtual assistants to content generation and more.
Let’s see some answers from ChatGPT-3.5 Fig. \ref{fig:Fig_1} when we ask “What are the basic maneuvers in driving a car?”\footnote{We intentionally cite an "older" ChatGPT  model to highlight the capabilities early on.}.
%

ChatGPT
\begin{itemize}
 \item {\it Driving in Traffic: Maintaining a safe following distance, staying in your lane, and being aware of other vehicles.} 
 \item {\it Merging: Safely merging onto highways and merging out of lanes or exit ramps.} 
 \item {\it Yielding Right of Way: Knowing when to yield to other vehicles, pedestrians, and traffic signs or signals.} 
 \item {\it Intersections: Navigating different types of intersections, including stop signs, yield signs, and traffic lights.} 
 \item {\it Etc.}
\end{itemize}

It will give similar responses when prompted about piloting an airplane or cruising a boat. Let’s see an example from following list, where we prompt ChatGPT 3.5 to write up Reqs for autonomous vehicles and further decompose them.

\begin{itemize}
 \item 
 User \\
 {\it We are developing an autonomous vehicle product. You are a systems engineer related to functional safety. Write me the functional requirements for urban driving.}
 \item ChatGPT \\
 {\it Obstacle Detection and Avoidance:}
 \begin{itemize}
\item {\it The autonomous vehicle shall detect and identify static and dynamic obstacles in its path during urban driving.}
\item And more on: {\it
Traffic Sign Recognition, Pedestrian Detection and Protection, Lane Keeping and Lane Changing, Intersection Handling, Emergency Maneuvers, …}
 \end{itemize}
 
 \item User \\
 {\it Further decompose the requirement that you gave me "The autonomous vehicle shall detect and identify static and dynamic obstacles in its path during urban driving" into more granular requirements.}
 \item ChatGPT: \\
 {\it
 Obstacle Identification:
 \begin{itemize}
\item {\it For static obstacles, the vehicle shall identify and classify objects such as buildings, road signs, and stationary vehicles}
\item And more on: {\it
Obstacle Detection, Obstacle Tracking, Obstacle Size and Shape Estimation, Sensor Fusion, Obstacle Database, False Positive/Negative Mitigation,…}
 \end{itemize}
 }
\end{itemize}

Arguably humans can decompose similar Reqs; possibly better, possibly worse (we have seen it both ways). However this was done over split seconds. There are two important observations to make. They are nicely written, and although high level, they are unambiguous. Thus we propose two applications: 1) Use LLMs for Reqs audit Section \ref{ReqsCheck} and 2) Functional decomposition Section \ref{FuncyD}.

\subsection{LLMs for requirements audit}
\label{ReqsCheck}
The idea is to use LLMs for Reqs audit. Parse human written Reqs and score them against objective criteria, with some being specificity, measurability, etc. Look at the following bullet list below. It is truly impressive! 
\begin{itemize}
 \item User 
 {\it \\Is the requirement for autonomous vehicles, "the autonomous vehicle would be fault operable" a nicely written requirement, if yes why, if no why?}
 \item ChatGPT
 {\it
 \\ The requirement "the autonomous vehicle would be fault operable" is not well-written and lacks clarity. A well-written requirement should be clear, unambiguous, and measurable. The phrase "fault operable" is vague and does not provide specific criteria or conditions for what is expected. Additionally, it does not specify how the system should behave in the presence of faults.
 Here are some reasons why the requirement is not well-written:
 \begin{itemize}
 \item Lack of Specificity:
 The term "fault operable" does not clearly define how the vehicle should operate in the presence of faults. It is essential to specify the desired behavior or conditions when faults occur.
 \item Measurability:
 The requirement is not easily measurable. It does not provide criteria or metrics for determining whether the vehicle meets the specified level of fault operability.
 \item Ambiguity:
 The term "fault operable" is open to interpretation. Different stakeholders may have different understandings of what it means for a vehicle to be fault operable.
 \item …
 \\ An improved version of the requirement could be:
 "The autonomous vehicle shall be designed to detect and respond to faults in a manner that ensures safe operation. In the event of a fault, the system shall implement appropriate fault mitigation strategies, including transitioning to a fail-safe mode, providing timely alerts to occupants, and initiating a safe stop if necessary. ..."
 \end{itemize}
 }
\end{itemize}

\subsection{Functional decomposition}
\label{FuncyD}
\begin{figure}[ht]
 \centering
 \includegraphics[width=\linewidth]{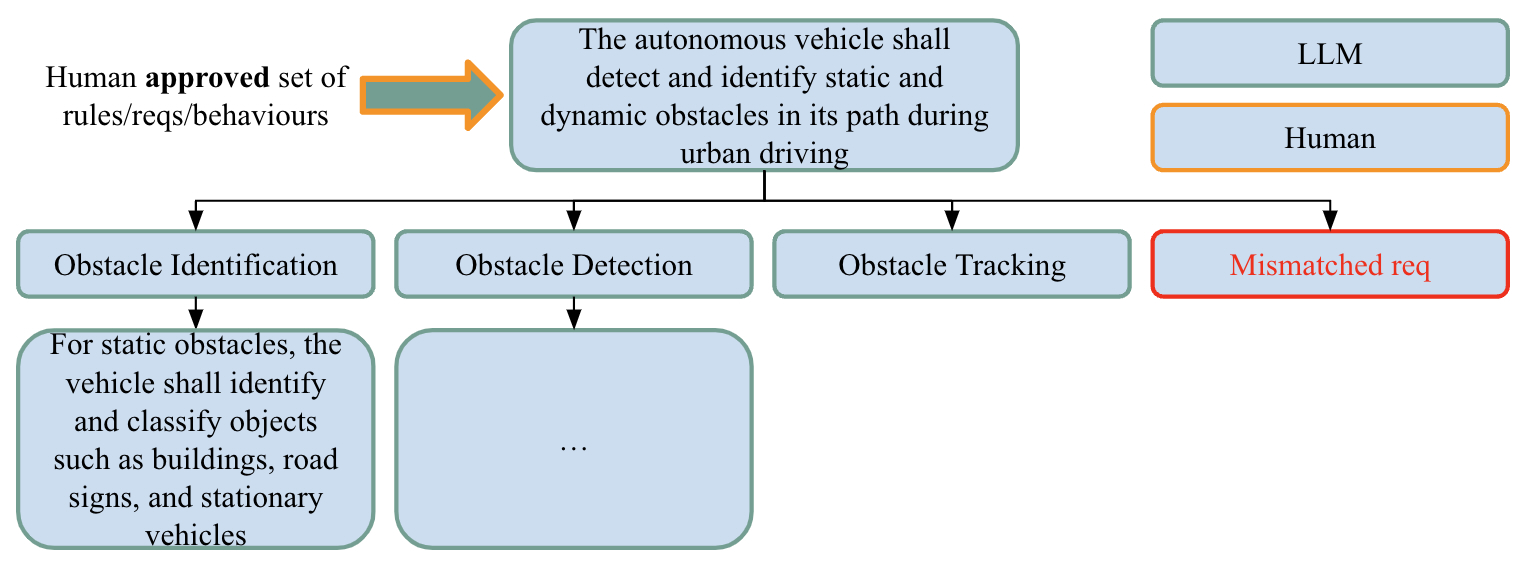} 
 \caption{LLM for functional decomposition and Reqs coverage assessment. The “Mismatched req” implies Reqs that have been identified as being different and/or missing from the human decomposed Reqs.}
 \label{fig:Fig_2}
\end{figure}

How can we tell that a system shall meet its design intent-objective; i.e. an AV being safe? A full Reqs functional decomposition could help us answer that: if all the children Reqs leading to a top level Req are met, then in a fully decomposed system, so is the parent. In case a Req is not met, it becomes a known risk that can be analyzed and addressed with possibly more development work or with certain mitigations that go beyond SW, such as limitations in the system etc. I.e. the AV system in a dense fog condition will slow down and maintain greater distances and/or gaps to other agents; i.e. \cite{28}. However, functional decomposition is more an art rather than a science and it is hard, a priori, to think of all the different permutations of the world-reality. Thus the idea is to use LLM for functional decomposition and Reqs coverage assessment. We shall create an API that assess human decomposed Reqs around parent-child relationships and traceability, as-well-as completeness. Such application would create two scores per Req and will flag humans for missing relationships (c.f. Fig. \ref{fig:Fig_2}). 

The parent-children relationships and functional decomposition can be done in various ways and is therefore easy to drift out. Therefore, the system shall start from a human approved Req and will assess the functional decomposition in a broader Req tree and/or different levels of the Reqs, using the appropriate tree searching technique to meet its objective. When it finds a satisfying match it will trace down the decomposition and match it to LLM generated one as complete or incomplete-missmatched. 

\begin{figure}[ht]
 \centering
 \includegraphics[width=\linewidth]{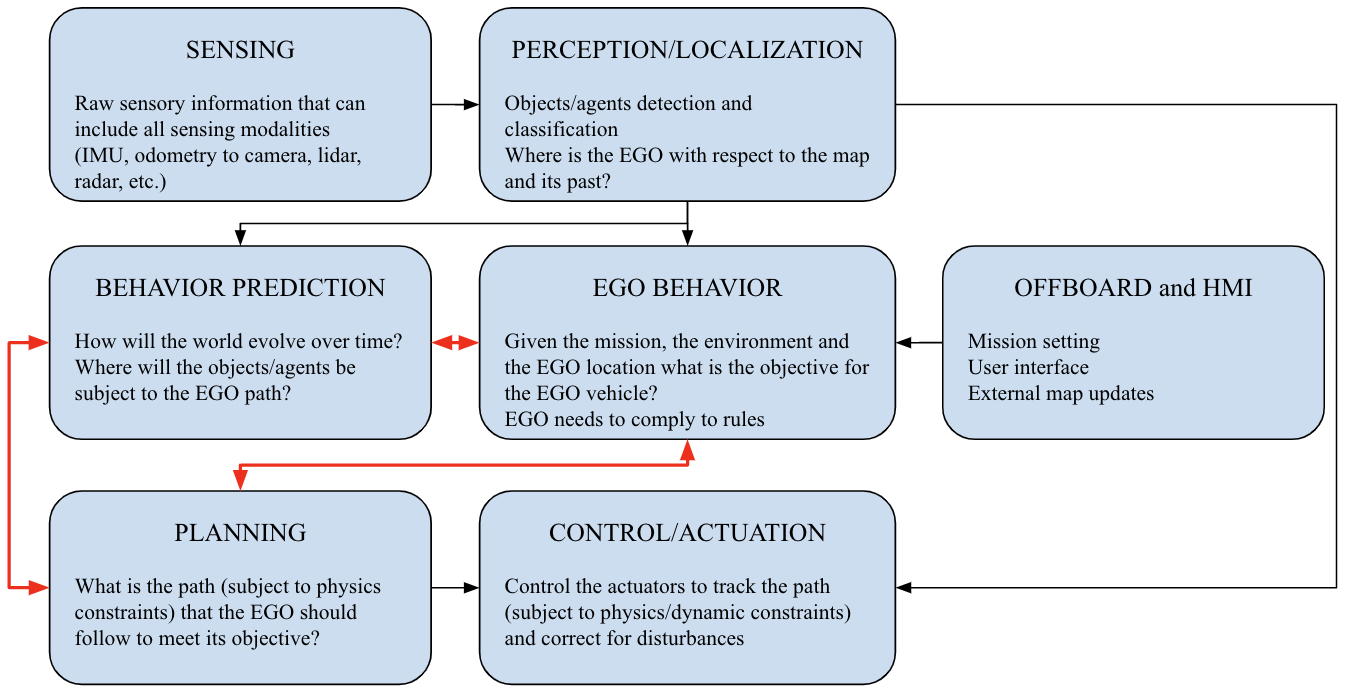} 
 \caption{The figure shows the Autonomous vehicle (AV) stack (abstraction from \cite{11,12,13,14}). EGO derives from the Greek word Εγω, that means myself. It refers to the AV itself as an entity in the AV navigation task. The three red arrow connectors between BEHAVIOR PREDICTION, EGO BEHAVIOR and PLANNING, depict that all three are depending on each other. Consider an AV leaving a zebra crossing while an occluded pedestrian (agent) appears in the scene. The pedestrian anticipates the AV to continue its path and would plan to pass behind the AV. If the AV comes suddenly to a halt (to avoid a probable collision), then it is possible that the pedestrian will stumble on the AV that was expected to be ahead (c.f. Explanation from Waymo \cite{15}).}
 \label{fig:Fig_3}
\end{figure}

\begin{figure}[ht]
 \centering
 \includegraphics[width=\linewidth]{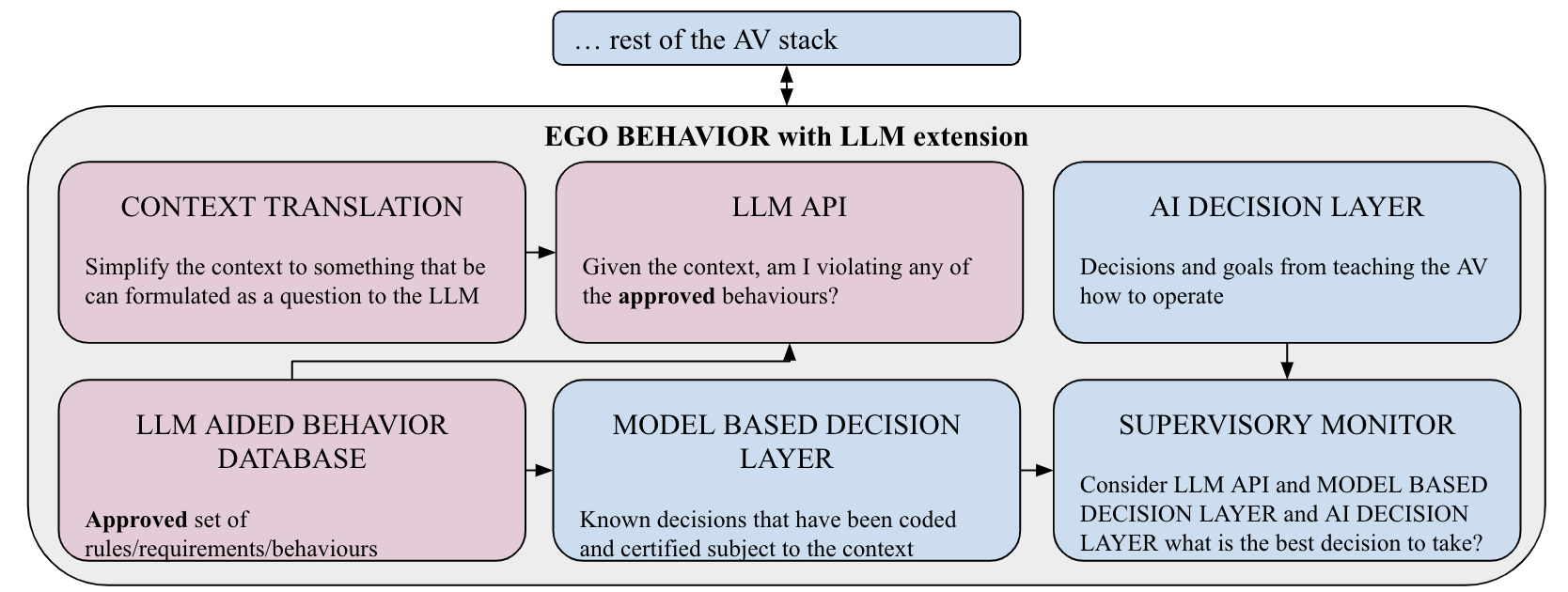} 
 \caption{LLM aided AV navigation using a supervisory monitor. The whole architecture focuses on the EGO BEHAVIOR of Fig. \ref{fig:Fig_3} however, the same concept can be applied in the other modules. The three red blocks depict the ones where most of the LLM functionality resides.}
 \label{fig:Fig_4}
\end{figure}

\section{AV Supervisory control}
\label{AVControl}

 Referring i) to Fig. \ref{fig:Fig_3} and using ii) the terminology of “Decision and Goal Making” from “The general architecture of the computational model for agents” from. J. Sifakis (c.f. \cite{11}), with the help of LLM we can:
\begin{enumerate}
 \item Generate (offboard) an approved behavioral set aided with LLM (LLM AIDED BEHAVIOR DATABASE) 
 \item Use a CONTEXT TRANSLATION and a LLM API to allow LLM to assess approved behaviors
 \item Generate an onboard SUPERVISORY MONITOR that improves the decision and goal making process
 \item Use the Fig. \ref{fig:Fig_4} architecture to create onboard and offboard metrics to assess the functional maturity on the development and verification and validation process
\end{enumerate}

We can “Generate (offboard) an approved behavioral set from LLM” as described in Section \ref{FuncyD}. 
We can therefore use LLMs to improve the explainability of the EGO BEHAVIOR decision making process. But why is explainability important? Primarily for the Safety Of The Intended Functionality (SOTIF)(c.f. \cite{16}). A concept invented for the usage of ML in AVs, where traditional functional safety standards (i.e. the famous ISO 26262 \cite{17}) cannot be applied. Refer to \cite{24} for a survey of explainability in various domains of Supervised ML. 

As written in the Reqs Development (completeness), the full functional decomposition is more of an art rather than a science. Therefore, the innovative architecture of Fig. \ref{fig:Fig_4} can be applied/used to enable the “safeguarding and supervisory systems” as written in the problem statement. It would serve as a “pillow” for behaviors and/or situations that may have not yet been seen in the field and/or, may have not yet caused problems.

When the AV behavior is monitored with diverse and/or different from a human Reqs decomposition lens it could flag inconsistencies during the verification phase. Similar concepts of supervisor monitors in SW systems that need high integrity are very common, acting as trusted testers. We aspire that such architecture can increase the “trustworthiness” highlighted so intensively by Professor Sifakis \cite{11} (Turing award winner in highly effective verification technology).

The main principle around this Section is to Create an LLM aided AV navigation using a supervisory monitor following steps a) to d).

\section{AV Supervisory control; proof-of-concept}
\label{AVControlPOF}
This Section is a proof-of-concept for the Section \ref{AVControl} AV Supervisory control. It is a simulation based on:
\begin{itemize}
 \item LLM model for Google Gemini. \url{https://ai.google.dev/api/models} using its available API and Google AI toolset \url{https://ai.google.dev/}\\
 \item Integrated in Google Colab: \url{https://colab.research.google.com/}\\
 \item A simple planar (bicycle) 3-DOF vehicle dynamics model with Euler integration at 10 ms
 \item Employs time varying Linear Quadratic Regulator (LQR) for lateral control (path tracking). Refer to Appendix Section \ref{Appendix} for the dynamics, control and specifics of the SW \footnote{In the SW, the "time.sleep(4)" controls the LLM requests rate. 15 requests per minute was feasible at no extra cost at the time of the simulations, Sept. 2024.}.  
\end{itemize}

\textbf{What is the simulation scenario?}

{\em A vehicle} is driving on the road abiding by the target speed limit. 
\vspace*{-5pt}
\begin{itemize}
 \item Two pedestrians are crossing the road without looking. They have their individual fixed speed: Crossing\_speed\_m\_s \\
 \vspace*{-0.2in}
 \item They start crossing the road {\em away} from the EGO at a: distance\_from\_EGOx0\_m
\\(c.f. Python class \textit{Pedestrians} below)
\end{itemize}

\begin{lstlisting}[]{python}
class Pedestrians:
 def __init__(self, name, distance_from_EGOx0_m,crossing_speed_m_s,
 lateralPosition_m,time2cross_s):
 self.name=name
 self.distance_from_EGOx0_m = distance_from_EGOx0_m
 self.crossing_speed_m_s = crossing_speed_m_s
 self.lateralPosition_m = lateralPosition_m
 self.time2cross_s = time2cross_s
\end{lstlisting}

\begin{figure}[ht]
 \centering
 \includegraphics[width=\linewidth]{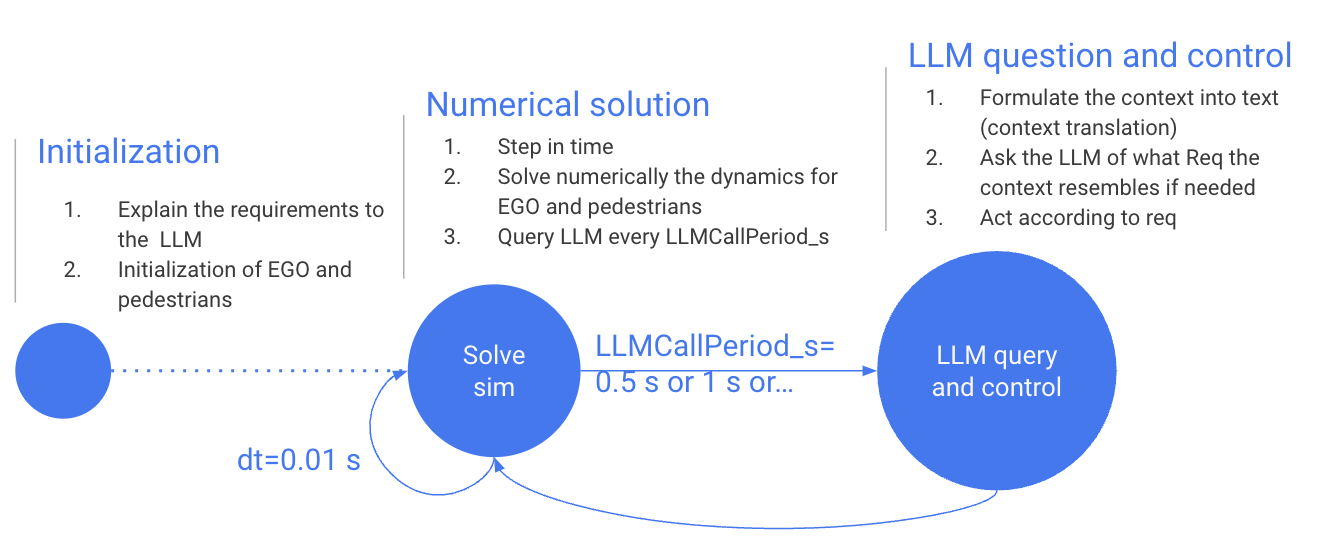} 
 \caption{SSuperLLM simulation process.}
 \label{fig:Fig_5}
\end{figure}

\subsection{Rules for the LLM}
\label{Rules}
Refer to Fig. \ref{fig:Fig_5}. We start by explaining the rules to the LLM, using the Google Gemini API with the prompt:
\\ \textit{You are an autonomous vehicle driver. The vehicle you are driving is called EGO. You need to drive subject to the following requirements:}
\begin{it}{
 \begin{enumerate}
 \item If a pedestrian is jaywalking along EGO's path AND is expected to have crossed the road by the time EGO will reach them, then the EGO shall maintain speed.
 \item If a pedestrian is jaywalking along EGO's path AND is expected to be close to crossing the road by the time EGO will reach them, and EGO and pedestrian are within soft braking distance, then the EGO shall apply soft braking and shall nudge away from the pedestrian.
 \item If a pedestrian is jaywalking along EGO's path AND is expected to be close to crossing the road by the time EGO will reach them, and EGO and pedestrian are within medium braking distance, then the EGO shall apply hard braking and shall nudge away from the pedestrian.
 \item If a pedestrian is jaywalking along EGO's path AND is expected to be in the middle of the road by the time EGO will reach them, and EGO and pedestrian are within soft braking distance, then the EGO shall apply medium braking.
 \item If a pedestrian is jaywalking along EGO's path AND is expected to be in the middle of the road by the time EGO will reach them, and EGO and pedestrian are within medium braking distance, then the EGO shall apply hard braking.
 \item If a pedestrian is jaywalking along EGO's path AND is expected to be on the road by the time EGO will reach them, and EGO and pedestrian are within hard braking distance, then the EGO shall apply full braking.
 \item If the EGO has slowed down below the route target speed OR nudged due to a pedestrian jaywalking AND there is no pedestrian on the road, then the EGO shall accelerate smoothly to the target route speed and shall go back to center of the road.
 \end{enumerate}
}\end{it}
We also explain the "LLM response" subject to the system state-conditions:
\begin{it}{
\begin{itemize}
 \item A requirement is met if all the AND conditions in the requirement are met.
 \item You need to respond with "Req=1, accel=0,nudge=0" if you encounter all the conditions for Requirement 1.
 \item You need to respond with "Req=2, accel=-2,nudge=1" if you encounter all the conditions for Requirement 2.
 \item You need to respond with "Req=3, accel=-4,nudge=1" if you encounter all the conditions for Requirement 3.
 \item You need to respond with "Req=4, accel=-4,nudge=0" if you encounter all the conditions for Requirement 4.
 \item You need to respond with "Req=5, accel=-6,nudge=0" if you encounter all the conditions for Requirement 5.
 \item You need to respond with "Req=6, accel=-8,nudge=0" if you encounter all the conditions for Requirement 6.
 \item You need to respond with "Req=7, accel=2,nudge=2" if you encounter all the conditions for Requirement 7.
 \item You can only respond with the text within the quotes "" but (not include the quote " symbol itself) and not come up with your answers because there is a program that parses your responses and is expecting a fixed structure.
\end{itemize}
}\end{it}

\subsection{Simulation results}
Section \ref{Appendix} explains in detail the vehicle dynamics and control (see Python code below). 
There is a common setup for both simulation examples shown in the paper, with only LLM call period (system decision period). 
\begin{lstlisting}[]{python}
## Initial Conditions
X_0 = 0# Initial x-position (m)
Y_0 = 2# Initial y-position (m)
Psi_0 = 0 # Initial yaw angle (rad)
x_dot0 = 10 # Initial longitudinal velocity (m/s)
## Pedestrian setup
obj1 = Pedestrians("Ped1", X_0+distanceToStopSoft_m+20,1.5,0,1)
obj2 = Pedestrians("Ped2", X_0+distanceToStopSoft_m+40,2,0,5)
\end{lstlisting}
\label{SimRes}
\subsubsection{Simulation with LLM call period at 0.5 s}
\label{LLM0p5s}
\begin{lstlisting}[]{python}
LLMCallPeriod_s=0.5
\end{lstlisting}

\begin{figure}[ht]
 \centering
 \includegraphics[width=\linewidth]{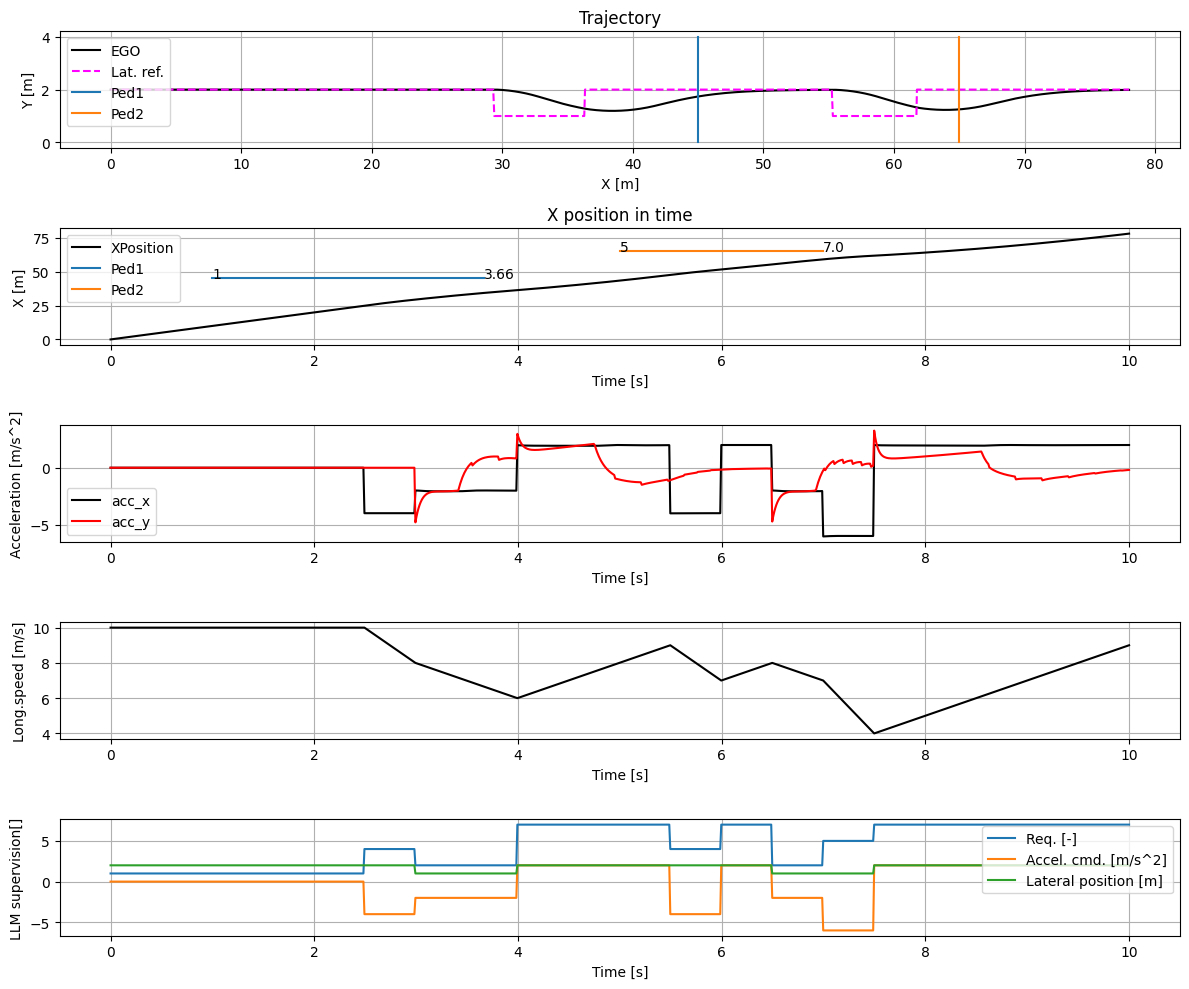} 
 
 \caption{Simulation with LLM call period at 0.5 s. The LLM Supervisory control successfully manages the task.}
 \label{fig:Fig_6}
\end{figure}

Refer to Fig. \ref{fig:Fig_6} for the LLMCallPeriod\_s=0.5s. The AV supervisory control successfully managed the communication and decision making using the CONTEXT TRANSLATION layer API depicted at Fig. \ref{fig:Fig_4}.

Below you can see how the communication between the API and LLM response looks like (c.f. Fig. \ref{fig:Fig_5}).\footnote{We only show some instances of the simulation.}
\begin{itemize}
 \item ...
 \item {\em State and Question}: EGO is traveling at 8.0 m/s I am 29.39 m longitudinally from X0 2.0 m laterally from Y0. Distance to Ped1 is 15.61 m with distanceToStopHard \footnote{
Travel distance to stop the vehicle with Soft, Medium, Hard and Very Hard Brake for -2, -4, -6 and -8 $m/s^2$ acceleration correspondingly.} 5.33 m. A pedestrian is jaywalking and is expected to be close crossing the road and EGO and pedestrian are within soft braking distance. What should the EGO do?
 \\ {\em LLM Response}: LLMTalks: "Req=2, accel=-2,nudge=1"
 \item {\em State and Question}: EGO is traveling at 8.0 m/s I am 29.39 m longitudinally from X0 2.0 m laterally from Y0. Distance to Ped2 is 9999 m with \footnote{"Distance to Ped* is 9999 m" means there is not a pedestrian crossing the road where *:=1 or 2.} distanceToStopHard 5.33 m. There is not a pedestrian along the EGO's path and EGO travels with less than target speed or has nudged. What should the EGO do?
 \\ {\em LLM Response}: LLMTalks: "Req=7, accel=2,nudge=2"
 \item {\em State and Question}: EGO is traveling at 7.0 m/s I am 33.12 m longitudinally from X0 1.68 m laterally from Y0. Distance to Ped1 is 11.88 m with distanceToStopHard 4.08 m. A pedestrian is jaywalking and is expected to be close crossing the road and EGO and pedestrian are within soft braking distance. What should the EGO do?
 \\ {\em LLM Response}: LLMTalks: "Req=2, accel=-2,nudge=1"
 \item {\em State and Question}: EGO is traveling at 7.0 m/s I am 33.12 m longitudinally from X0 1.68 m laterally from Y0. Distance to Ped2 is 9999 m with distanceToStopHard 4.08 m. There is not a pedestrian along the EGO's path and EGO travels with less than target speed or has nudged. What should the EGO do?
 \\ {\em LLM Response}: LLMTalks: "Req=7, accel=2,nudge=2"
 \item ...
\end{itemize}

\subsubsection{LLM call period at 2 s}
\label{LLM2s}
\begin{lstlisting}[]{python}
LLMCallPeriod_s=2
\end{lstlisting}

Refer to Fig. \ref{fig:Fig_6} for the LLMCallPeriod\_s=2s. The 2s is too big of time gap to take decisions. The "Oups!" shows a virtual spatial overlap between vehicle and pedestrian.

\begin{figure}[ht]
 \centering
 \includegraphics[width=\linewidth]{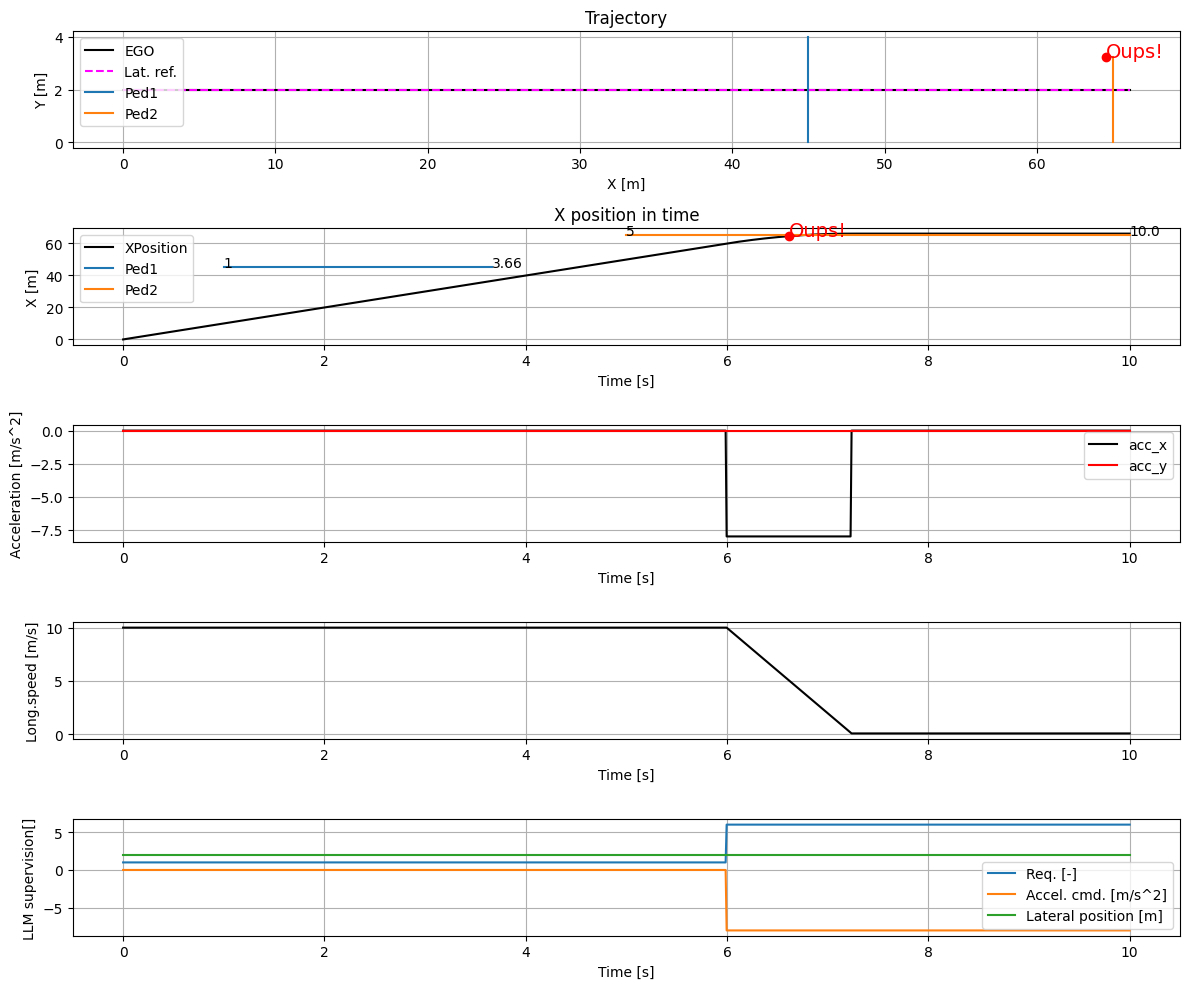} 
 
 \caption{LLM call period at 2 s. As expected, the LLM Supervisory control due to very low supervision frequency is not successful to manage the task.}
 \label{fig:Fig_7}
\end{figure}

\subsubsection{Does “Supervisory control” work?}
\label{Solved}
Reviewing the results of the simulations from Fig. \ref{fig:Fig_6} and Fig. \ref{fig:Fig_7} it appears that the concept is working. However the reality is that we are on right path but not done yet. 

Let’s see another (older) formulation of Reqs-rules that we used during the research phase. 

\begin{it}{
 \begin{enumerate}
 \item ...
 \item If a pedestrian is jaywalking along EGO's path AND is expected to be \textbf{close crossing the road} by the time EGO will reach them, and the distance between EGO and pedestrian is \textbf{less than a soft braking distance, then the EGO shall apply soft braking and shall nudge 1 meter away from the pedestrian}.
 You need to respond with "Req.2, accel=-2,nudge=yes" if you encounter the condition for Requirement 2.
 \item If a pedestrian is jaywalking along EGO's path AND is expected to \textbf{in the middle of the road} by the time EGO will reach them, and the distance between EGO and pedestrian is less than a soft braking distance,then the EGO \textbf{shall apply medium braking}.
 You need to respond with "Req.3, accel=-4,nudge=no" if you encounter the condition for Requirement 3.
 \item If a pedestrian is jaywalking along EGO's path AND is expected to \textbf{in the middle of the road}by the time EGO will reach them, and the distance between EGO and pedestrian \textbf{is less than a medium braking distance, then the EGO shall apply hard braking}.
 You need to respond with "Req.4, accel=-6,nudge=no" if you encounter the condition for Requirement 4.
 \item ...
 \end{enumerate}
}\end{it}

Below you can see the how communication between the API and LLM response looks like for the above Reqs formulation.
\begin{itemize}
 \item {\em State and Question}: A pedestrian is jaywalking and is expected to be close crossing the road and the distance between EGO and pedestrian is less than medium braking. What should the EGO do?
 \\ {\em LLM Response}: LLMTalks: "Req.3, accel=-4,nudge=no"
\end{itemize}
The above seems wrong\footnote{This result was taken on April 2024. LLMs and AI evolve very fast so is highly likely that the output shall not be the same when re-run.}. \textbf{But why?}
We are asking the LLM to reason between close crossing the road and in the middle of the road. We have not given explicit instructions… a human is hypothesized to reason nudge and hard braking. 

\section{Discussion}
\begin{figure}[ht]
 \centering
 \includegraphics[width=\linewidth]{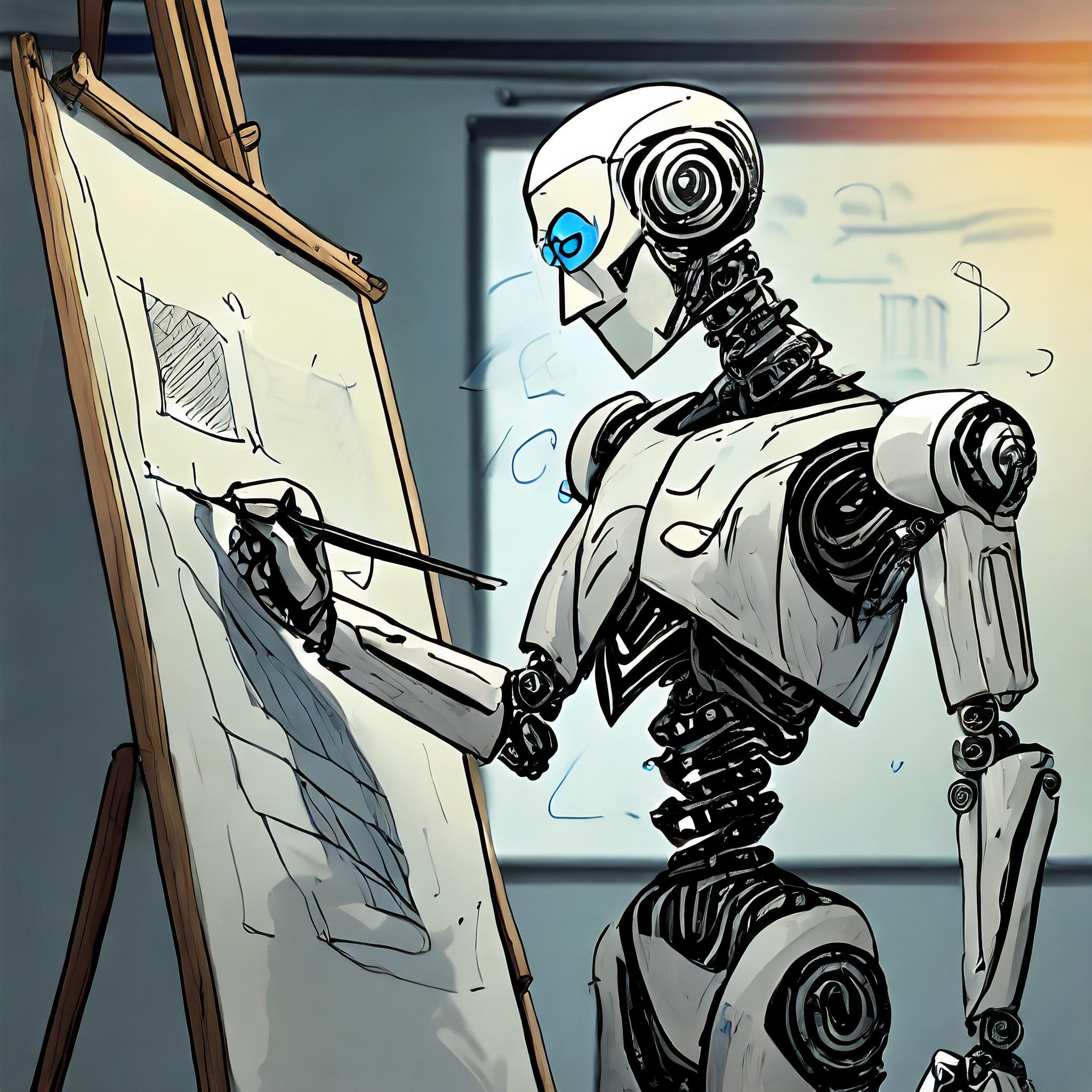} 
 \caption{The System engineer of the future as imagined by AI.}
 \label{fig:Fig_8}
\end{figure}

The proof-of-concept shows that our hypothesis for LLM and SysEng is true; however more research is due and caution is needed before drawing definitive conclusions. Humans need to be involved (still) and rules are needed: a) the maturity of the LLM model influences the results and b) the requirements (Reqs) should be written explicitly without anticipating any reasoning. Reqs have to be: atomic and self contained, clear, measurable, unambiguous, non-overlapping, etc. GAI and LLMs can increase someone’s productivity by orders of magnitude. Still AI is a productivity multiplier. Should System Engineers, System Architects and SW programmers be worried about the future of their work? We are at the cusp of a new order of things with AI (c.f. Fig. \ref{fig:Fig_8}). We need to embrace the change and go with it. 

\section*{Acknowledgments}
The author is grateful to Professor Sifakis \cite{11} for their insightful and inspiring discussions about Systems Engineering and Trustworthiness of Autonomous Systems. The author is also grateful to postgraduate student Marios Konstantoudakis. 

\section{Appendix. Dynamics and Control}
\label{Appendix}
\subsection{Vehicle dynamics}
The non-linear dynamical equations of the planar (bicycle) 3-DOF vehicle dynamics model used in Section \ref{AVControlPOF} are in equations (1) to (7) (c.f. \cite{22}). For more in depth fundamentals of vehicle dynamics the reader is refered to \cite{29}.

\setcounter{equation}{0}

\begin{align}
\alpha_f &= \delta - \arctan \left( \frac{\dot{y} + \dot{\psi} l_f}{\dot{x}} \right) \\
\alpha_r &= -\arctan \left( \frac{\dot{y} - \dot{\psi} l_r}{\dot{x}} \right) \\
F_{yf} &= C_f \alpha_f \\
F_{yr} &= C_r \alpha_r \\
\ddot{x} &= a_{xCmd} \\
\ddot{y} &= \frac{1}{m} \left( F_{yf} \cos(\delta) + F_{yr} \right) - \dot{\psi} \dot{x} \\
\ddot{\psi} &= \frac{1}{I_z} \left( l_f F_{yf} - l_r F_{yr} \right)
\end{align}

Where
$\alpha_f$ and $\alpha_r$ are the front and rear wheel slip angles, respectively, $\dot{x}$, $\dot{y}$ and $\dot{\psi}$ are longitudinal, lateral and yaw velocity, $F_{yf}$ and $F_{yr}$ are the front and rear lateral forces, respectively, and the road wheel angle $\delta$ and longitudinal acceleration command $a_{xCmd}$ are the longitudinal and lateral the control commands respectively. The parameters are shown in Table \ref{tab:vehicle_params}.

\begin{table}[ht]
\centering
\caption{Vehicle Parameters}
\begin{tabular}{|l|c|}
\hline
Parameter & Value \\
\hline
$m$ : Mass (kg) & 1470 \\
$I_z$ : Moment of Inertia (kg*$m^2$) & 1900 \\
$l_f$ : Front Axle Distance (m) & 1.04 \\
$l_r$ : Rear Axle Distance (m) & 1.56 \\
$C_f$ : Front Tire Cornering Stiffness (N/rad) & 71000 \\
$C_r$ : Rear Tire Cornering Stiffness (N/rad) & 47000 \\
\hline
\end{tabular}
\label{tab:vehicle_params}
\end{table}

\subsection{Control}

Control is split between longitudinal and lateral control. 

\textbf{Longitudinal control}

The output of the Supervisory control determines the acceleration command (c.f. Section \ref{Rules}). This acceleration command via Euler integration becomes the vehicle's speed (c.f. equation (5)).

\textbf{Lateral control}

\begin{figure}[ht]
 \centering
 \includegraphics[width=\linewidth]{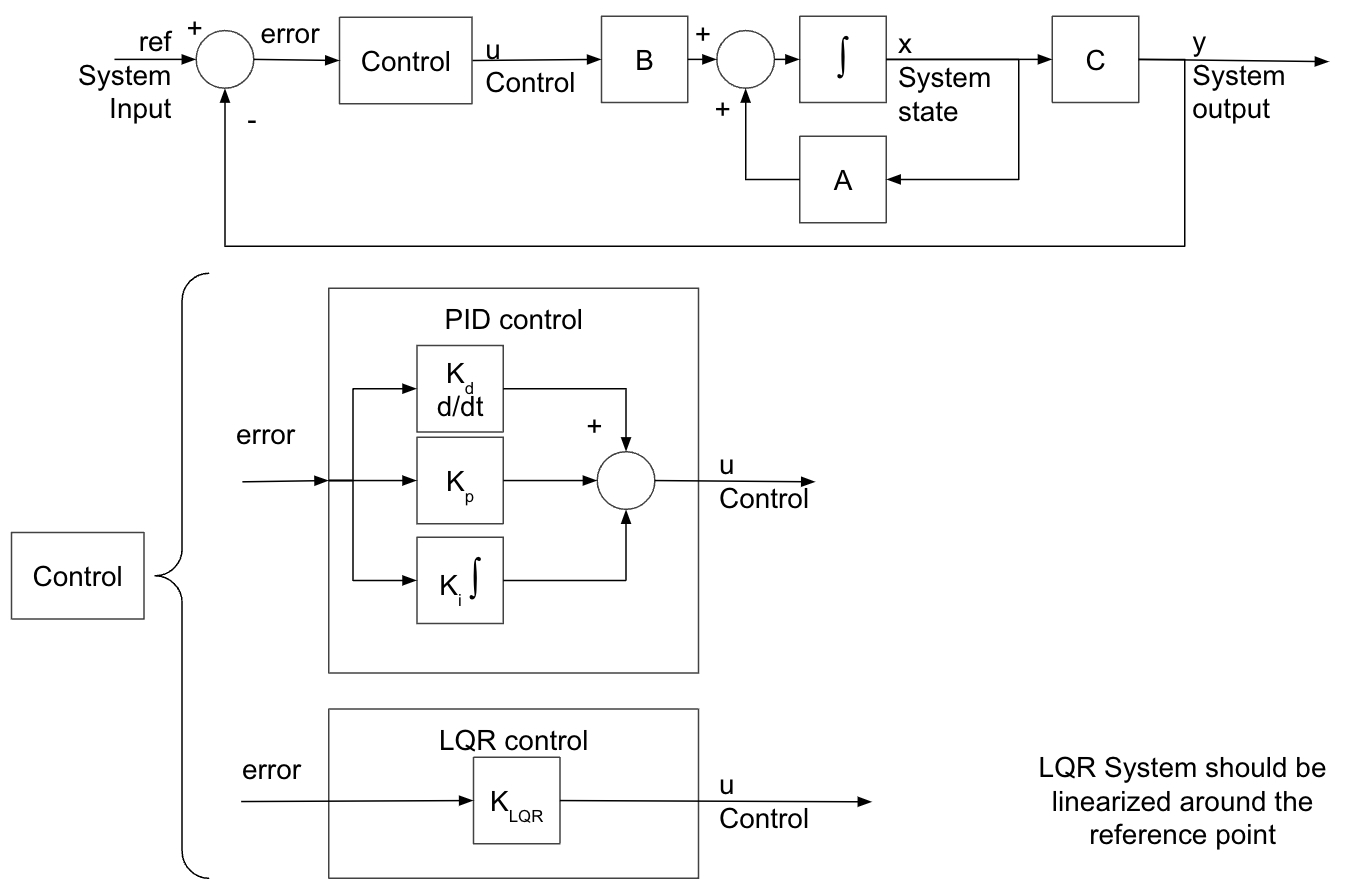} 
 \caption{LQR vs PID block diagram. $A$, $B$ and $C$ are the State, Input-to-state and State-to-output matrices correspondingly. The "System Input" is the reference(ref), where for Section \ref{SimRes} are the vehicle's lateral ($Y$) and yaw position ($\Psi$) (c.f. equation (8)).} 
 \label{fig:Fig_9}
\end{figure}

The system is controlled via an LQR (c.f. \cite{23}) using a linearized state-space version \eqref{2} of the non-linear equations (1) to (7). Refer to Fig. \ref{fig:Fig_9} for visual representation of the LQR against a common PID controller. The \emph{updateLQR} function (see below) is using the continuous-time state-space matrices A and B from \eqref{2}, solves the associated algebraic Riccati equation $P_{cont}$ and calculates correspondingly the optimal gain matrix $K_{LQR}$. For simplicity we use the Python function "solve\_continuous\_are" \footnote{
State cost matrix $Q$ and Input cost matrix $R$ are configured for lateral tracking as: Q = np.array([[0.5,0,0,0],[0,0.3,0,0],[0,0.0,0.0,0],[0,0,0,0.3]]), R = np.array([[5]]). The setup penalizes deviation from the reference states but also reduces abrupt changes into the control input $\delta$.} to solve the Riccati \footnote{An alternative approach is to solve the Riccati as a boundary condition problem with numerical integration as described here \url{https://www.mathworks.com/matlabcentral/answers/94722-how-can-i-solve-the-matrix-riccati-differential-equation-within-matlab}}. The \emph{updateLQR} is executed every time there is a $\Delta$ speed (speed difference) greater than 0.5 m/s (configurable). This allows for optimal lateral control with changing vehicle's speed.

\begin{lstlisting}[]{python}
def updateLQR(x_dot):
 A = np.array([
 [0, 1, x_dot,0],
 [0,-(C_f + C_r) / (mass * x_dot),0, 
 -(C_f * l_f - C_r * l_r) / (mass * x_dot) - x_dot],
 [0,0,0,1],
 [0,-(C_f * l_f - C_r * l_r) / (I_z * x_dot), 0, 
 -(C_f * l_f**2 + C_r * l_r**2) / (I_z * x_dot)]
 ])
 P_cont = solve_continuous_are(A, B, Q, R)
 K_LQR_cont = np.linalg.inv(R) @ (B.T @ P_cont)
 K_LQR=K_LQR_cont
 return K_LQR_cont
\end{lstlisting}

\begin{equation}
\label{2}
\frac{d}{dt}
\begin{bmatrix}
Y\\
\dot{y} \\
\Psi\\
\dot{\psi}\\
\end{bmatrix}=\\
\begin{bmatrix}
0 & 1 & \dot{x} & 0 \\
0 & \frac{-(C_f+C_r)}{m\dot{x}} & 0 & \frac{-l_f\cdot C_f+C_r\cdot l_r}{m\dot{x}} -\dot{x} \\
0 & 0 & 0 & 1 \\
0 & \frac{-l_f\cdot C_f+C_r\cdot l_r}{I_z\dot{x}} & 0 & \frac{-({l_f}^2\cdot C_f+C_r\cdot {l_r}^2)}{I_z\dot{x}}
\end{bmatrix}\cdot \begin{bmatrix}
Y\\
\dot{y} \\
\Psi\\
\dot{\psi}\\
\end{bmatrix}+\begin{bmatrix}
0\\
\frac{C_f}{m} \\
0\\
\frac{l_f\cdot C_f}{I_z}\\
\end{bmatrix}\cdot\delta\\
\end{equation}

\vskip 0.2in

\bibliographystyle{unsrt}  
\bibliography{references}  

\begin{thebibliography}{10}

\bibitem{1}
Nick Webb, Dan Smith, Christopher Ludwick, Trent Victor, Qi~Hommes, Francesca Favaro, George Ivanov, and Tom Daniel.
\newblock Waymo's safety methodologies and safety readiness determinations.
\newblock {\em arXiv preprint arXiv:2011.00054}, 2020.

\bibitem{2}
Kristofer~D Kusano, John~M Scanlon, Mattias Br{\"a}nnstr{\"o}m, Johan Engstr{\"o}m, and Trent Victor.
\newblock Framework for a conflict typology including contributing factors for use in ads safety evaluation.
\newblock In {\em 27th International Technical Conference on the Enhanced Safety of Vehicles (ESV) National Highway Traffic Safety Administration}, number 23-0328, 2023.

\bibitem{4}
Eli Bronstein, Mark Palatucci, Dominik Notz, Brandyn White, Alex Kuefler, Yiren Lu, Supratik Paul, Payam Nikdel, Paul Mougin, Hongge Chen, et~al.
\newblock Hierarchical model-based imitation learning for planning in autonomous driving.
\newblock In {\em 2022 IEEE/RSJ International Conference on Intelligent Robots and Systems (IROS)}, pages 8652--8659. IEEE, 2022.

\bibitem{5}
B~Varadarajan, A~Hefny, A~Srivastava, KS~Refaat, N~Nayakanti, A~Cornman, K~Chen, B~Douillard, CP~Lam, D~Anguelov, et~al.
\newblock Efficient information fusion and trajectory aggregation for behavior prediction.
\newblock {\em URL https://arxiv. org/abs/2111.14973}, 2021.

\bibitem{6}
Yeping Hu, Wei Zhan, and Masayoshi Tomizuka.
\newblock Probabilistic prediction of vehicle semantic intention and motion.
\newblock In {\em 2018 IEEE Intelligent Vehicles Symposium (IV)}, pages 307--313. IEEE, 2018.

\bibitem{7}
Eli Bronstein, Sirish Srinivasan, Supratik Paul, Aman Sinha, Matthew O’Kelly, Payam Nikdel, and Shimon Whiteson.
\newblock Embedding synthetic off-policy experience for autonomous driving via zero-shot curricula.
\newblock In {\em Conference on Robot Learning}, pages 188--198. PMLR, 2023.

\bibitem{8}
Federal Highway~Administration US~Department~of Transportation.
\newblock Systems engineering for its handbook - section 3 what is systems engineering?, 2023.

\bibitem{26}
Marius Bozga and Joseph Sifakis.
\newblock Safe by design autonomous driving systems.
\newblock {\em arXiv preprint arXiv:2405.11995}, 2024.

\bibitem{0}
Georgios Gatos, Spyridon Karakostas, Andreas Agiotis, and Diomidis Katzourakis.
\newblock Vehicle dynamics and suspension design using systems engineering.
\newblock In {\em 2024 32nd Mediterranean Conference on Control and Automation (MED)}, pages 328--333. IEEE, 2024.

\bibitem{25}
Anthony Corso, Robert Moss, Mark Koren, Ritchie Lee, and Mykel Kochenderfer.
\newblock A survey of algorithms for black-box safety validation of cyber-physical systems.
\newblock {\em Journal of Artificial Intelligence Research}, 72:377--428, 2021.

\bibitem{27}
Joseph Sifakis and David Harel.
\newblock Trustworthy autonomous system development.
\newblock {\em ACM Transactions on Embedded Computing Systems}, 22(3):1--24, 2023.

\bibitem{9}
Long Chen, Oleg Sinavski, Jan H{\"u}nermann, Alice Karnsund, Andrew~James Willmott, Danny Birch, Daniel Maund, and Jamie Shotton.
\newblock Driving with llms: Fusing object-level vector modality for explainable autonomous driving.
\newblock In {\em 2024 IEEE International Conference on Robotics and Automation (ICRA)}, pages 14093--14100. IEEE, 2024.

\bibitem{18}
Chetan Arora, John Grundy, and Mohamed Abdelrazek.
\newblock Advancing requirements engineering through generative ai: Assessing the role of llms.
\newblock In {\em Generative AI for Effective Software Development}, pages 129--148. Springer, 2024.

\bibitem{19}
Markus Borg.
\newblock Requirements engineering and large language models: Insights from a panel.
\newblock {\em IEEE Software}, 41(2):6--10, 2024.

\bibitem{20}
Nick Feng, Lina Marsso, Sinem~Getir Yaman, Isobel Standen, Yesugen Baatartogtokh, Reem Ayad, Vict{\'o}ria~Oldemburgo de~Mello, Beverley Townsend, Hanne Bartels, Ana Cavalcanti, et~al.
\newblock Normative requirements operationalization with large language models.
\newblock In {\em 2024 IEEE 32nd International Requirements Engineering Conference (RE)}, pages 129--141. IEEE, 2024.

\bibitem{21}
Mohamad Fazelnia, Viktoria Koscinski, Spencer Herzog, and Mehdi Mirakhorli.
\newblock Lessons from the use of natural language inference (nli) in requirements engineering tasks.
\newblock {\em arXiv preprint arXiv:2405.05135}, 2024.

\bibitem{10}
I~SAE.
\newblock Sae levels of driving automation™ refined for clarity and international audience.
\newblock {\em https://www.sae.org/blog/sae-j3016-update}, 2021.

\bibitem{28}
Diomidis Katzourakis, John~P Alsterda, Rami~Y Hindiyeh, Robert Chen, and Carson Schultz.
\newblock Motion planning constraints for autonomous vehicles, January~25 2024.
\newblock US Patent App. 17/894,812.

\bibitem{11}
Joseph Sifakis.
\newblock Autonomous systems--an architectural characterization.
\newblock {\em Models, Languages, and Tools for Concurrent and Distributed Programming: Essays Dedicated to Rocco De Nicola on the Occasion of His 65th Birthday}, pages 388--410, 2019.

\bibitem{12}
David Gonz{\'a}lez, Joshu{\'e} P{\'e}rez, Vicente Milan{\'e}s, and Fawzi Nashashibi.
\newblock A review of motion planning techniques for automated vehicles.
\newblock {\em IEEE Transactions on intelligent transportation systems}, 17(4):1135--1145, 2015.

\bibitem{13}
Diomidis Katzourakis, Christopher~L Porritt, Johannes~A Huennekens, Huibert Mees, and Paul~J Keas.
\newblock Active suspension system with energy storage device, October~27 2020.
\newblock US Patent 10,814,690.

\bibitem{14}
Diomidis Katzourakis, Huibert Mees, and Paul~W Choin.
\newblock Steer-by-wire system with multiple steering actuators, September~29 2020.
\newblock US Patent 10,787,192.

\bibitem{15}
Yuxiao Chen, Sushant Veer, Peter Karkus, and Marco Pavone.
\newblock Interactive motion planning for autonomous vehicles with joint optimization.
\newblock {\em arXiv preprint arXiv:2310.18301}, 2023.

\bibitem{16}
Shaoshan Liu.
\newblock Council post: Safety of the intended functionality (sotif) for autonomous driving.
\newblock {\em https://www.forbes.com/sites/forbestechcouncil/2022/09/23/safety-of-the-intended-functionality-sotif-for-autonomous-driving/}, 2022.

\bibitem{17}
ISO.
\newblock Iso 26262-1:2018, road vehicles — functional safety, part 1: Vocabulary.
\newblock {\em https://www.iso.org/standard/68383.html}, 2018.

\bibitem{24}
Nadia Burkart and Marco~F Huber.
\newblock A survey on the explainability of supervised machine learning.
\newblock {\em Journal of Artificial Intelligence Research}, 70:245--317, 2021.

\bibitem{22}
Diomidis Katzourakis, Joost~CF de~Winter, Stefan de~Groot, and Riender Happee.
\newblock Driving simulator parameterization using double-lane change steering metrics as recorded on five modern cars.
\newblock {\em Simulation Modelling Practice and Theory}, 26:96--112, 2012.

\bibitem{29}
Rajesh Rajamani.
\newblock {\em Vehicle dynamics and control}.
\newblock Springer Science \& Business Media, 2011.

\bibitem{23}
Dimitri Bertsekas.
\newblock {\em Dynamic programming and optimal control: Volume I}, volume~4.
\newblock Athena scientific, 2012.

\end{thebibliography}

\end{document}